\documentclass{icrc2009}

\usepackage{graphicx}   
\usepackage{caption}    
\usepackage{fixltx2e}
\usepackage{url}

\newcommand{\shorttitle}[1]%
{\markboth{Proceedings of the 31\MakeLowercase{$^{st}$} ICRC, {\L}\'{o}d\'{z} 2009}{#1} }
\newcommand{\etal}{\MakeLowercase{\textit{et al. }}} 
\def\Journal#1#2#3#4{{#1} {\bf #2}, #3 (#4)}

\begin{document}
\title{Atmospheric MUons from PArametric formulas:\\a fast GEnerator
for neutrino telescopes (MUPAGE)}

\author{\IEEEauthorblockN{M. Bazzotti\IEEEauthorrefmark{1}, S. Biagi\IEEEauthorrefmark{1}\IEEEauthorrefmark{2},
    G. Carminati\IEEEauthorrefmark{1}\IEEEauthorrefmark{2}, S. Cecchini\IEEEauthorrefmark{1}\IEEEauthorrefmark{3},
    \\T. Chiarusi\IEEEauthorrefmark{2}, A. Margiotta\IEEEauthorrefmark{1}\IEEEauthorrefmark{2}, M.
    Sioli\IEEEauthorrefmark{1}\IEEEauthorrefmark{2} and M. Spurio\IEEEauthorrefmark{1}\IEEEauthorrefmark{2}
                  }
                            \\
\IEEEauthorblockA{\IEEEauthorrefmark{1}Dipartimento di Fisica
dell'Universit\`a di Bologna, Viale Berti Pichat 6/2, 40127 Bologna,
Italy} \IEEEauthorblockA{\IEEEauthorrefmark{2} INFN, Sezione di
Bologna, Viale Berti Pichat 6/2, 40127 Bologna, Italy}
\IEEEauthorblockA{\IEEEauthorrefmark{3}INAF-IASF, Via Gobetti 101,
40129 Bologna, Italy}}

\shorttitle{G. Carminati \etal MUPAGE generator} \maketitle

\begin{abstract}
Neutrino telescopes are opening new opportunities in observational
high energy astrophysics. In these detectors, atmospheric muons from
primary cosmic ray interactions in the atmosphere play an important
role. They provide the most abundant source of events for
calibration and for testing the reconstruction algorithms. On the
other hand, they represent the major background source.

The simulation of a statistically significant number of muons in
large volume neutrino telescopes requires a big effort in terms of
computing time. Some parameterizations are currently available, but
they do not explicitly take into account the arrival of muons in
bundles. A fast Monte Carlo generator (MUPAGE) was developed to
generate single and multiple atmospheric muon events in
underwater/ice neutrino telescopes. The code reduces the computing
time for the simulation of atmospheric muons significantly.

The event kinematics is produced on the surface of a user-defined
cylinder, virtually surrounding the detector volume. The flux of
muon bundles at different depths and zenith angles, the lateral
spread and the energy spectrum of the muons in the bundles are based
on parametric formulas~\cite{paper,maurizio}. These formulas were
obtained according to a specific primary cosmic ray flux model and
constrained by the measurements of the muon flux in the MACRO
underground experiment~\cite{macro}.

Some MUPAGE applications are presented.
  \end{abstract}

\begin{IEEEkeywords}
 Simulation of atmospheric muons, neutrino telescopes, Monte Carlo
 generator
\end{IEEEkeywords}

\section{Introduction}
The most abundant signals for a neutrino telescope are due to high
energy muons resulting from the extensive air showers produced by
interactions between primary cosmic rays (CRs) and atmospheric
nuclei. Although neutrino telescopes are located at large depths
under water or in ice, taking advantage of the shielding effect
offered by these detector media, a large flux of high energy
atmospheric muons can reach the active volume of the detectors.

The atmospheric muons represent an insidious background for track
reconstruction as their Cherenkov light can mimic fake upward going
tracks. This kind of signatures can be confused with the cosmic
neutrino signals that neutrino telescopes are searching for. On the
other hand, atmospheric muons are a useful tool to test offline
analysis software, to check the understanding of the detector and to
estimate systematic uncertainties. Moreover, the pointing capability
of the telescopes can be studied using atmospheric muons through the
measurement of the moon shadow.

Atmospheric muon bundles can be accurately reproduced by a full
Monte Carlo (MC) simulation (e.g. CORSIKA~\cite{corsika}), starting
from primary CR interactions with atmospheric nuclei, generating the
resulting large number of air showers and propagating the muons
through sea water or ice. A full MC requires a large amount of
computing time and therefore the production of a large statistical
event sample cannot be easily obtained for a detector of a cubic
kilometer size. In a recent search for a diffuse flux of neutrinos
\cite{icecubediffuse}, the use of a full MC simulation limited the
number of generated atmospheric muons at 63 equivalent detector days
to represent a background for 807 active days of data.

In order to save the big effort of computing time, a fast Monte
Carlo generator is therefore essential. Some parameterizations for
the underwater/ice flux and energy spectrum are available in
literature~\cite{okada,bugaev,klimu}. None of them, however, gives
the possibility to simulate two or more muons produced in the same
CR interaction (muon bundle). In this paper the event generator
MUPAGE is presented. It is based on parametric formulas and takes
into account the multiplicity of muon bundles.

\section{Simulation of atmospheric muon bundles}

Parametric formulas are derived in \cite{paper,maurizio} and
describe the flux, the angular distribution and the energy spectrum
of underwater/ice muon bundles. The muon flux and energy spectrum
are parameterized, for the first time, in terms of the bundle
multiplicity $m$.

\begin{figure}[!ht]
\centering \vspace{-3mm}
\includegraphics[width=18pc]{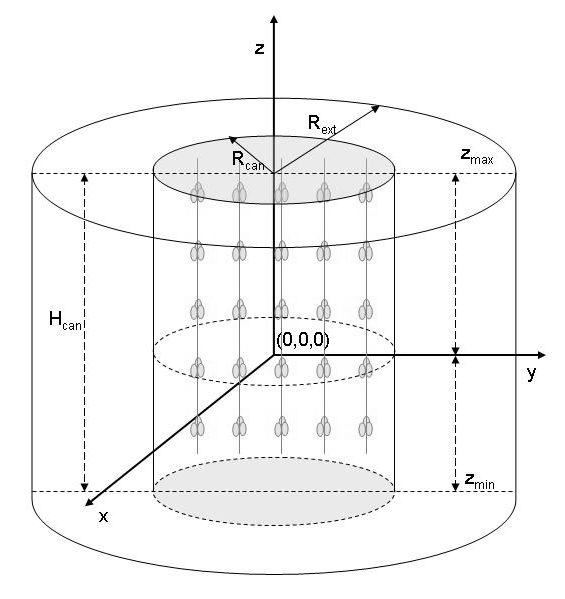}
\vspace{-4mm} \caption{\small{Sketch of some input parameters. The
cylinder surrounding the instrumental volume is the {\it can}, with
radius $R_{\mathrm{can}}$ and height $H_{\mathrm{can}}$. The events
are generated on an extended can with $R_{\mathrm{ext}}$. The origin
of the coordinate system does not have to be located at the center
of the detector. The lower disk is at a depth $H_{\mathrm{max}}$
with respect to the sea/ice surface.}} \label{fig:cylinder}
\end{figure}

From these parametric formulas an event generator called MUPAGE
(MUon GEnerator from PArametric formulas)~\cite{mupage} was
developed. The program output is a formatted ASCII table containing
the kinematics of atmospheric muon bundles on the surface of a {\it
can}. The {\it can} (Fig. \ref{fig:cylinder}) is an imaginary
cylinder surrounding the active volume of any proposed detector
between 1.5 and 5.0 km of water equivalent depth. The output table
with the event kinematics can be used as input for the following
steps of the detector-dependent MC simulation, which include the
production of Cherenkov light in water or ice and the simulation of
the signal in the detection devices as photomultiplier tubes.

The code is available in a {\it tar} archive~\cite{mupage} or by
sending a request to the authors.

\section{MUPAGE structure}
The MUPAGE code is written in C++ and it has been tested with gcc
version 3.2.x., 3.4.x and 4.1.x. The program requires ROOT
libraries~\cite{root} for the pseudorandom number generator. The
{\it tar} archive  contains the code, the \texttt{Makefile}, a
\texttt{README} file, two template script files (for tcsh and for
bash) to launch the executable and some subdirectories. After
compilation, the executable and a  \texttt{Linux/} directory,
containing the object files, are created. The core classes are
contained in directories \texttt{inc/} and \texttt{src/}, which do
not need to be changed by the user.

The simplest way to execute MUPAGE is to use, for tcsh, the C-shell
script file \texttt{run-mupage.csh}, or, for bash, the Shell script
file \texttt{run-mupage.sh}. Here, the user can modify  the random
seed, the run number and the number of events to be generated
$N_{gen}$.

\begin{figure}[!ht]
\centering \vspace*{-1mm}
\includegraphics[width=18pc]{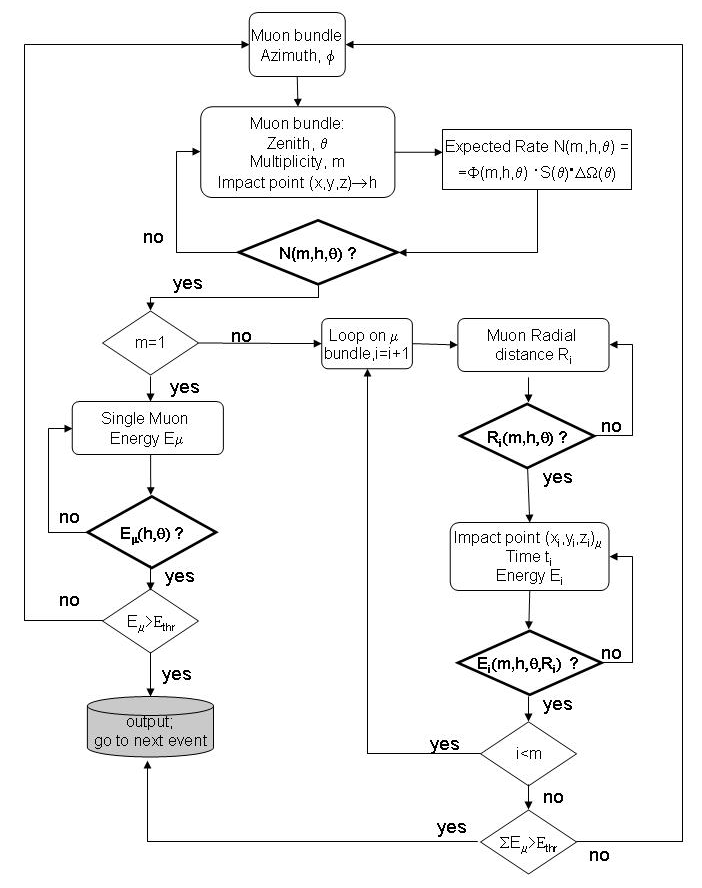}
\vspace{-2mm}\caption{\small{The flowchart of the MUPAGE event
generator. The smooth-angle rectangles indicate the extraction of
uniformly distributed random values. The decisional rhombuses in
bold select values according to formulas reported in~\cite{paper},
with a Hit-or-Miss method. The procedure is iterated for $N_{gen}$
events.}} \label{fig:flowchart}
\end{figure}

The flowchart of the Monte Carlo program is shown in Fig.
\ref{fig:flowchart}. The bundle multiplicity, direction and impact
point of the shower axis on the {\it can} surface are generated
first. According to the depth-zenith angle-intensity relation, this
event can be accepted or rejected, using a standard Hit-or-Miss
method. Once selected, if the event is a single muon, the energy is
sampled from the energy distribution expected for that specific
depth and zenith angle. If the event is a multiple muon, for each
muon the radial distance $R$ from the shower axis is sampled,
followed by the muon energy (which depends also on $R$). The impact
point of each muon in the bundle is the projection on the {\it can}
surface of its position on the plane perpendicular to the shower
axis, where it has been generated. Consequently, since all muons in
the bundle are assumed to reach that plane at the same time, also
the arrival time of each muon on the {\it can} surface is computed.
Details are given in~\cite{mupage}. Since the multiplicity $m$ of
the bundle is generated according to the muon flux, it can happen
that some muons in the bundle do not geometrically intercept the
{\it can} surface and therefore they are not written in the output
file.

\vspace*{-4.5mm}

\subsection{Description of the input file}\label{ssec:input}
The generator needs some parameters as input: the dimensions and
depth of the {\it can} surrounding the detector (see Fig.
\ref{fig:cylinder}), the density of the detector medium and the
ranges of the various simulation parameters (e.g. multiplicity,
zenith angle, muon energy). The user can change the default values
(which refer to a production done for the ANTARES
experiment~\cite{moriond}). The main parameters are:

\noindent\textbf{Hmax} (2.475 km): vertical height of the sea/ice
level with respect to the {\it can} lower disk;

\noindent \textbf{Zmin} ($-278.15$ m): position of the {\it can}
lower disk  in the detector frame, \textbf{Zmax} (313.97 m):
position of the {\it can} upper  disk. The {\it can} height is
defined as: $H_{\mathrm{can}} = Z_{\mathrm{max}} -
Z_{\mathrm{min}}$;

\noindent \textbf{Rcan} (238.61 m): {\it can} radius;
\textbf{EnlargedCanR} (100 m): {\it can} radius extension;

\noindent \textbf{density} (1.03 g cm$^{-3}$): mean value of the
density of the detector medium (ice or water) to convert
\textbf{Hmax} in km.w.e.;

\noindent \textbf{THETAmin} ($0^\circ$), \textbf{THETAmax}
($85^\circ$): minimum and maximum value of the zenith angle range;

\noindent \textbf{MULTmin} (1), \textbf{MULTmax} (100):  minimum and
maximum value of  bundle multiplicity range;

\noindent \textbf{Rmax} (100 m): maximum muon lateral spread with
respect to the shower axis;

\noindent \textbf{Ethreshold} (0.001 TeV): the minimum energy
threshold of all muons combined in the event bundle.

The radius of the generation volume is increased by the quantity
\textbf{EnlargedCanR} in order to accept also muons very far from
the shower axis. Since one of the aims of the code is to save
computing time, the user is encouraged to set the same value for
\textbf{EnlargedCanR} and \textbf{Rmax}.

An example of usage of parameter \textbf{Ethreshold} is given in
Sect.~\ref{ssec:antares}.

\subsection{Description of the output files}\label{ssec:output}
The output file (\textbf{\$OUTFILE1} in the template script)
contains all information about the generated events in a formatted
ASCII table. An example of an output data file with 1000 generated
events (\texttt{run\_01.evt}) can be found in the {\it tar} archive.
Each line of the table contains the following information:

\noindent \textbf{evt\_id m $track\_id$ $x_{i}\ y_{i}\ z_{i}\ v_x\
v_y\ v_z\ E_{i}\ t_{i}\ G\_id$}

\noindent where:
\begin{itemize}
\item \textbf{evt\_id} is the event number;
\item \textbf{m} is the multiplicity of the muon bundle at the depth
where the shower axis hits the {\it can};
\item for each muon $m_c$ in the bundle intercepting the {\it can}
$(m_c\le m)$:
\begin{itemize}
\item[-] $track\_id = i$ ($i = 1, m_c$) is the muon identifier in
the event;
\item[-] $(x_i, y_i, z_i) $ are the coordinates [in m] of the muon
impact point on the {\it can} surface;
\item[-] $(v_x, v_y, v_z)$ are the direction cosines of the muon,
coincident with those of the bundle axis;
\item[-] $E_i$ is the energy [in GeV] of the muon;
\item[-] $t_i$ is the time delay [in ns] of the $i$-th muon at the
{\it can} surface with respect to the first muon $(i=1)$. $t_i$ can
be either positive or negative;
\item[-] $G\_id$ is the GEANT particle identification number (6 =
$\mu^-$ by default).
\end{itemize}
\end{itemize}

The livetime and its statistical error (see
Sect.~\ref{ssec:livetime}) corresponding to the total number of
simulated events $N_{gen}$ is written in a second file
(\textbf{\$OUTFILE2}).

Note that there is no bias or ordering (in energy, multiplicity and
zenith angle) in the simulation. All events have the same weight so
the output file reproduces a real data file. A disadvantage of this
approach is that since all the input physics parameters (primary CR
flux, interaction models, muon transport) are fixed, the events
cannot be re-weighted.

\subsection{Equivalent livetime of the simulation} \label{ssec:livetime}

From $N_{gen}$, the equivalent livetime of the run is automatically
calculated. Since the flux of muon bundles of a given multiplicity
$m$ is known for each value of depth $H$ at a given zenith angle
$\vartheta_i$, the  expected rate on the \textit{can} upper disk at
the depth $H=H_{\mathrm{min}}$ is:

\begin{displaymath}
\dot{N}_{MC}( \Delta \Omega_i)={\Phi(H_{\mathrm{min}},\vartheta_i ,
m) \cdot S \cdot \Delta \Omega_i} \quad [\mathrm{s}^{-1}]
\end{displaymath}

\noindent where, $\Delta \Omega_i = 2 \pi (\cos\vartheta_{1i} -
\cos\vartheta_{2i})$ is the  solid angle centered at
$\vartheta_i=(\vartheta_{1i}+\vartheta_{2i})/2$, and $S = \pi
R_{ext}^2 \cos\vartheta_i \ $ [m$^2$] is the projected area of the
upper disk. The equivalent livetime is computed from the number of
generated events $N(\Delta \Omega_i )$ on the upper disk in the
solid angle $\Delta \Omega_i$ as:

\begin{displaymath}
T(\Delta \Omega_i) =N(\Delta \Omega_i )/\dot{N}_{MC}(\Delta
\Omega_i) \quad [\mathrm{s}] \label{eq:livetime}
\end{displaymath}

The livetime is computed as the weighted average on 33 different
solid angle regions of $T(\Delta \Omega_i)$, which have the same
value, within statistical errors, and are written in the
\textbf{\$OUTFILE2} file.

\section{Examples and application} \label{sec:examples}

The development of MUPAGE was motivated by the need of a large
sample of atmospheric muons in order to simulate the response and
the possible background for the neutrino studies in the ANTARES and
NEMO Experiments and moreover to study possible detector geometries
for the future cubic kilometer detector in the Mediterranean Sea, on
behalf of the KM3NeT Consortium. Although a full Monte Carlo
simulation is also available~\cite{brunner}, MUPAGE offers the
possibility of a faster simulation. Some examples of the MUPAGE
performance are presented.

\subsection{Usage in the ANTARES Collaboration}\label{ssec:antares}
The ANTARES neutrino telescope~\cite{moriond} has been taking data
since March 2006 and it has been completed in May 2008. During the
first 6 months of operation, in its one line configuration,
measurements of atmospheric muons from data are taken. For this
first analysis, two samples of atmospheric muon bundles with
different multiplicity ranges are generated using MUPAGE: for $m = 1
- 30$, the $8.8 \times 10^8$ generated events correspond to a
detector livetime of 11.7 days; and for $m = 31 - 1000$, the $8
\times 10^6$ generated events correspond to 32.4 equivalent detector
days. After the MUPAGE kinematics generation, muons are propagated
inside the {\it can}, where the emission of Cherenkov light and the
production of signals in the photomultiplier tubes (PMTs) are
simulated with a software package~\cite{brunner} which needs a
computing time almost a factor 10 larger than that required by
MUPAGE. The two samples are used to investigate the efficiency of
the track reconstruction method, reproducing the time and the
angular residuals very well, and to measure the vertical muon
intensity versus depth~\cite{line1}, in good agreement with other
published values. In particular for the latter goal, the MUPAGE
simulation was used to convert the measured rate of reconstructed
tracks to the single muon intensity.

For the 5 line configuration of the ANTARES detector, in 2007,
MUPAGE is used to generate a data set with livetime of one month.
This production is split into 322 files ($<$ 2 GB each). Each file
contains $10^7$ events, corresponding to $8060\pm 3$ s. The
computing time needed to produce a file (on a 2xIntel Xeon Quad
Core, 2.33 GHz per core) is less than 2 hours. This MUPAGE data set
is used to compare the zenith and azimuth angle distributions of
reconstructed tracks with the real data and with a CORSIKA full
Monte Carlo simulation~\cite{annarita}. A preliminary comparison
between data and MC of electromagnetic showers produced by high
momentum muons is also performed~\cite{mangano}. The MUPAGE
simulation enables the development of a deconvolution method which
to be applied to the real data in order to obtain the experimental
atmospheric muon flux as a function of the sea depth~\cite{marco}.

The same MC data set is additionally used to estimate the background
rate induced by simultaneous muon bundles originating from different
cosmic rays. If two atmospheric muons arrive during a trigger
window, the produced signals on the PMTs cannot be distinguished
from each other. In particular, the timing patterns of the light
from the two tracks can be such that the reconstruction result is a
single upward going track. The rate of triggering `coupled events',
which reach the 5 line ANTARES detector within a time window smaller
than 4 $\mu$s, is about $7\times 10^{-4}$~Hz. However, no events are
reconstructed as upward going, fixing the 90\% confidence level
upper limit at $9 \times 10^{-7}$~Hz.

A third data set is produced for the background study of the high
energy neutrino diffuse flux (above 100~TeV). In order to optimize
the computing time, a (conservative) cut on the minimum energy
threshold of all muons combined in the bundle is applied, $E_{thr}>$
3~TeV, and the event multiplicity is divided in several ranges,
split is 852 files. The total computing time is 232 hours with a
livetime of one year.

\subsection{Usage in the NEMO Collaboration}
The NEMO Collaboration~\cite{nemo} has looked for an optimal site
for the installation as well as for the development of technologies
for an underwater cubic kilometer experiment. The NEMO Phase-1
project, operating from December 2006 to May 2007, has allowed a
first validation of the feasibility of cubic kilometer
detector~\cite{tonino}. A small MUPAGE MC data set is produced,
generating $4 \times 10^7$ events corresponding to $\sim 11.3$
equivalent hours. The MC simulation of zenith and azimuth
distribution of reconstructed tracks is used for the comparison with
11.3 hours of data taking, giving an excellent
agreement~\cite{amore}.

\subsection{Usage in the KM3NeT Consortium}
The KM3NeT Consortium aims at the definition of a complete project
for an underwater cubic kilometer neutrino telescope to be installed
in the Mediterranean Sea~\cite{km3net}. Since the final detector
performance will depend critically on the rejection of atmospheric
muons, MC simulations for different detector geometries as a
function of site depth are performed. For example, a small MUPAGE MC
data set with a livetime of 105~min is used to study the response to
atmospheric muons for different PMT orientations~\cite{piera}.

\section{Conclusion}

MUPAGE is a fast generator of the kinematics of atmospheric muon
bundles. It uses parametric formulas, valid in the interval $1.5 \le
h \le 5.0$ \ km water equivalent and $\vartheta \le 85^\circ$, for
the flux of single and, for the first time, multiple muons. The
simulation of the energy spectrum of single and multiple muons takes
into account the dependence of the muon energy on the shower
multiplicity and the distance of the muon from the shower axis. The
generator, developed to minimize the computing time, represents a
useful tool for underwater/ice neutrino telescopes for the
production of a large amount of simulated data. The code is
currently used for the Monte Carlo simulation of atmospheric muons
on behalf of the ANTARES and NEMO Collaborations. Furthermore, the
KM3NeT Consortium is using the code to study the atmospheric muon
rejection performances of different detector geometries.


\begin{thebibliography}{99}

\bibitem{paper}Y. Becherini {\it et al.}, \Journal{\em{Astrop.
Phys.}}{25}{1}{2006}.

\bibitem{maurizio} M. Spurio {\it et al.}, these proceedings.

\bibitem{macro} M. Ambrosio {\it et al.}, MACRO Collaboration,
\Journal{\em{Phys. Rev.} D}{56}{1407}{1997}; \Journal{\em{Phys.
Rev.} D}{56}{1418}{1997}.

\bibitem{corsika} D. Heck {\it et al.}, Report FZKA6019
(1998), Forshungszentrum Karlsruhe.

\bibitem{icecubediffuse} A. Achterberg {\it et al.}, \Journal{\em{Phys.
Rev.} D}{76}{042008}{2007}.

\bibitem{okada}  A. Okada, \Journal{\em{Astrop.
Phys.}}{2}{393}{1994}.

\bibitem{bugaev} E.V. Bugaev {\it et al.}, \Journal{\em{Phys.
Rev.} D}{58}{054001}{1998}.

\bibitem{klimu} S.I. Klimushin {\it et al.}, \Journal{\em{Phys.
Rev.} D}{64}{014016}{2001}.

\bibitem{mupage} G. Carminati {\it et al.}, \Journal{\em{Comput. Phys.
Commun.}}{179}{915}{2008}; arXiv:0802.0562~[physics.ind-det] (2008).

\bibitem{root} R. Brun and F. Rademakers, \Journal{{\em Nucl. Instrum.
Methods} A}{389}{81}{1997}; see also http://root.cern.ch

\bibitem{moriond} G. Carminati (for the ANTARES Coll.), in:
Proceedings of the Rencontres de Moriond 2009 EW session, La Thuile.
arXiv:0905.1373~[astro-ph] (2009).

\bibitem{brunner} J. Brunner (Antares simulation tools), in:
Proceedings of the VLVnT Workshop, October 5-8, 2003, Amsterdam.
http://www.vlvnt.nl/

\bibitem{line1} M. Ageron {\it et al.}, ANTARES Collaboration,
\Journal{\em{Astrop. Phys.}}{31}{227}{2009}.

\bibitem{annarita} A. Margiotta (for the ANTARES Coll.),
\Journal{{\em Nucl. Instrum. Methods} A}{602}{76}{2008}.

\bibitem{mangano} S. Mangano (for the ANTARES Coll.), \Journal{{\em
Nucl. Instrum. Methods} A}{588}{107}{2008}.

\bibitem{marco} M. Bazzotti (for the ANTARES Coll.), these
proceedings.

\bibitem{nemo} E. Migneco et al., NEMO Collaboration, \Journal{{\em Nucl.
Instrum. Methods} A}{567}{444}{2006}. See also:
http://nemoweb.lns.infn.it/

\bibitem{tonino} A. Capone {\it et al.} (NEMO Collaboration),
\Journal{{\em Nucl. Instrum. Methods} A}{602}{47}{2009}.

\bibitem{amore} I. Amore (for the NEMO Coll.), \Journal{{\em Nucl.
Instrum. Methods} A}{602}{68}{2009}.

\bibitem{km3net} KM3NeT Consortium: http://www.km3net.org

\bibitem{piera} P. Sapienza {\it et al.}, \Journal{{\em Nucl.
Instrum. Methods} A}{602}{101}{2009}.

\end{thebibliography}
\end{document}